\documentclass{aa}
\usepackage{graphicx}
\begin{document}
   \title{Astrometric Microlensing of Quasars}

\subtitle{Dependence on surface mass density and external shear}

   \author{Marie Treyer\inst{1}
          \and
	Joachim Wambsganss\inst{2}
          }

%   \offprints{J. Wambsganss}

   	\institute{
 		Laboratoire d'Astrophysique de Marseille, 
		Traverse du Siphon, 
		13376 Marseille, 
		France;
              \email{marie.treyer@oamp.fr}
         \and
		Universit\"at Potsdam, 
        		Institut f\"ur Physik, 
        		Am Neuen Palais 10, 
        		14467 Potsdam, 
        		Germany;
             \email{jkw@astro.physik.uni-potsdam.de}
             }

   \date{Received 5 September 2003; accepted 9 November 2003}

   \abstract{ 
A small fraction of all quasars are strongly lensed
and multiply imaged, with usually a galaxy acting as the main lens.
Some, maybe all of these quasars are also affected by microlensing, 
the effects of stellar mass objects in the lensing galaxy. 
Stellar microlensing not only has photometric effects 
(the apparent magnitudes of the quasar images vary independently 
due to the relative motion between source, lens and observer),
it also affects the observed position of the images. 
This astrometric effect was first explored by  Lewis and Ibata (1998): 
the position of the quasar -- i.e. the center-of-light of the many
 microimages -- can shift by  tens of microarcseconds due to the relatively 
sudden (dis-)appearance of a pair of microimages when a caustic is being crossed.

We explore this effect quantitatively for different values of the 
lensing parameters  $\kappa$ and  $\gamma$
(surface mass density and external shear)
covering most of the known multiple quasar systems.  
We show examples of microlens-induced quasar motion and the 
corresponding light curves for different quasar sizes. 
We evaluate statistically the occurrence of large 
shifts in angular position and their correlation with
apparent brightness fluctuations. 
We also show statistical relations between positional offsets and
time from random starting points.
As  the amplitude of the astrometric offset
depends on the source size, 
astrometric microlensing signatures of quasars 
-- combined with the photometric variations --
will provide very good constraints on the size of quasars as
a function of wavelength.
We predict that such signatures will be detectable 
for realistic microlensing scenarios 
with near future technology in the infrared/optical
(Keck-Interferometry, VLTI, SIM, GAIA).
Such detections will show that not even high
redshift quasars define a ``fixed'' coordinate system. 
   }

   \maketitle
%
%________________________________________________________________

\section{Introduction}
Gravitational lensing acts on quasars in a number of ways. The
best known effect is     strong lensing which       produces
multiply imaged quasars.
Only a small
fraction of all quasars (about 1 in 500)            
are multiply imaged 
with galaxies acting as the main lenses
(for an updated list, see
the CASTLES web page {\tt http://cfa-www.harvard.edu/glensdata/}).
Most of these quasars 
-- and possibly some ``single'' quasars as well --
are also affected by microlensing: the coherent effects of 
stellar mass objects in the lensing galaxy. 
Microlensing is well known 
for affecting the apparent brightness of the quasar,
which changes due to the relative  motion between lens, 
source and observer
(see, e.g., 
	Paczy\'nski 1986,
	Wambsganss, Paczy\'nski \& Schneider 1990, 
	Lewis et al. 1998,
	Wyithe et al. 2000a, 
	Wambsganss 2001, 
	Wo\'zniak et al. 2000a,b). 
However, stellar microlensing has yet another effect, 
which was first explored
by  Lewis and Ibata (1998): 
when the quasar crosses a (micro-)caustic, a new
pair of microimages is created or destroyed. 
The relatively sudden (dis-)appearance of
     a highly magnified image pair does  not just produce a strong
fluctuation 
                  in apparent brightness,     it
can also shift the center-of-light of the quasar 
(the weighted sum of all the microimages)
by       tens of microarcseconds. 
Such a positional change is in principle detectable
with current VLBI techniques in the radio regime, and will be
measurable
in the infrared and optical wavebands as well
with (near) future 
technology (Keck-Interferometry, VLTI, SIM, GAIA).

A number of studies have explored 
astrometric microlensing in a different regime,
namely stars in the Milky Way
(or dark matter objects in its
halo) acting on background stars either in the Magellanic
Clouds or in the bulge of the Milky Way 
(e.g., 
Miyamoto \& Yoshii 1995;
Miralda-Escude 1996;
Mao \& Witt 1998;
Boden, Shao \& van Buren 1998;
Goldberg \& Wo\'zniak 1998;
Paczy\'nski 1998;
Han, Chun \& Chang 1999;
Han \& Kim 1999;
Han \& Jeong 1999;
Safizadeh, Dalal, Griest 1999;
Dominik \& Sahu 2000;
Gould \& Han 2000;
Han \& Kim 2000;
Salim \& Gould 2000;
Delplancke et al. 2001; 
Belokurov \& Evans 2002;
Dalal \& Lane 2003).
The effects of Milky Way stars on background quasars were
studied by 
Hosokawa, Ohnishi \&  Fukushima (1997),
Sazhin et al. (1998) and Honma \& Kurayama (2002).
Williams \& Saha (1995) had
discussed large image shifts produced by substructure in
the lensing galaxy.
However, 
not much work has yet been done on cosmological 
astrometric microlensing -- stars in lensing galaxies acting on even
more distant quasars -- beyond Lewis \& Ibata (1998), with the recent
exception of Salata \& Zhdanov (2003). 
The present paper aims to explore this further.

We study quantitatively eight different cases with 
various   values of the
        surface mass density,  with and without external shear.
We present example microlensing lightcurves with the
corresponding center-of-light shifts and  we
investigate the correlation  
between high-magnification photometric events
and large-offset astrometric events. 

As first pointed out by Lewis \& Ibata (1998),
large offset events are   typically     
correlated with high-magnification
events, whereas the inverse is much less true. 
The reason for this is that 
the location of the newly appearing bright image pair  
during 
a caustic crossing is unrelated to
the  previous  center-of-light of the microimages.
In rare cases, 
the new image pair may appear at or close to 
the center-of-light of the pre-existing  microimages. 
Such a situation would correspond
to a large change in brightness with very little change in position.
In most cases, however, the new bright image pair will appear at a location
which is unrelated to the previous center-of-light, and hence
produce a sudden jump.
Since the positional offset is preferentially
perpendicular to
the direction of the external shear, the shear direction 
can be inferred this way.

After introducing the microlensing length and time scales
(Section 2), we describe our simulations (Section 3). 
In Section 4, 
we  illustrate the effect of astrometric and photometric
fluctuations and present 
statistical correlations between positional offsets, 
magnitude fluctuations and time intervals between the measurements
for four microlensing 
situations with different surface mass densities
($\kappa = 0.2, 0.4, 0.6, 0.8$), with and without 
external shear ($\gamma = \kappa$ or  0).
Finally  we discuss the possibilities  of real detections of this
phenomenon in the near future.

\section{Microlensing basics: length,  time and angular scales}

%*************
% table 1
%***********************************************************************
   \begin{table*}{}
      \centering
      \caption[]{Summary of scales for the 
      ``typical'' lensing case 
      ($z_L \approx 0.5$, $z_Q \approx 2$)
      and for the special case of Q2237+0305
      ($z_L \approx 0.039$, $z_Q \approx 1.69$):
      }
         \label{tab-cases}
         \begin{tabular}{lll}
	 \noalign{\smallskip}
        	& typical lensed quasar case 	& lensed quasar Q2237+0305 \\
	 \hline
Physical Einstein radius in quasar plane $R_E$   	& $3.2 \times 10^{16} \sqrt{M / M_\odot} ~\rm h_{75}^{-0.5}$ cm   

		& $1.6 \times 10^{16} \sqrt{M / M_\odot} ~\rm h_{75}^{-0.5}$ cm \\ 
Angular Einstein radius $\theta_E$	& $2.2 \sqrt {M /M_\odot} ~\rm h_{75}^{-0.5}~\mu$arcsec
		& $7.3 \sqrt {M /M_\odot} ~\rm h_{75}^{-0.5}~\mu$arcsec \\
Einstein time $t_E$		& $25  \sqrt {M / M_\odot} ~v_{600}^{-1}~\rm h_{75}^{-0.5}$ years
		& $8.7 \sqrt {M / M_\odot} ~v_{600}^{-1}~\rm h_{75}^{-0.5}$ years \\
Crossing time $t_{cross}$	& $210~R_{15} ~v_{600}^{-1} ~\rm h_{75}^{-0.5}$ days
		& $18~R_{15} ~v_{600}^{-1} ~\rm h_{75}^{-0.5}$ days \\
	 \noalign{\smallskip}            
	 \hline
         \end{tabular}
   \end{table*}
%********************************************************************

\subsection{Standard mass, length and time scales}
The lensing effects of  cosmologically distant 
compact objects in the mass range 
$  10^{-3} \le M/M_{\odot} \le 10^3$
on background sources is usually called ``cosmological microlensing".
The source is typically a distant quasar, but in principle other
        objects can be microlensed as well, 
e.g. high-redshift supernovae (e.g., Rauch 1991)
or
gamma-ray sources/bursters (Torres et al. 2003; 
Koopmans \& Wambsganss 2001).  
The only condition is that the source size
be comparable to or smaller than the Einstein radius of the 
foreground lens.

The microlenses can be ordinary stars, 
brown dwarfs, planets, black holes,
molecular clouds, or other compact mass concentrations.
In most practical cases, the microlenses are part of a galaxy which 
acts as the main (macro-)lens.  However, microlenses could also
be located in, say, clusters of galaxies 
(Tadros, Warren \& Hewett 1998,
Totani 2003)
or they could even be 
imagined ``floating" freely and filling intergalactic space
(Hawkins 1996, Hawkins \& Taylor 1997).

The relevant length scale for microlensing is the 
Einstein radius in the source plane:
	
\begin{equation}
 R_E = 
\sqrt{ { {4 G M } \over {c^2} } { {D_S D_{LS} \over D_L}  } } 
\approx  3.2 \times 10^{16} \sqrt{M \over M_\odot} ~\rm h_{75}^{-0.5}~cm
\end{equation}
where ``typical" lens and source redshifts of $z_L = 0.5$  and
$z_S = 2.0$  
were assumed for the expression
on the right hand side ($G$ and $c$ are the gravitational
constant and the velocity of light, respectively; $M$ is the mass of
the lens, $D_L$, $D_S$, and $D_{LS}$ are the angular diameter distances
between observer 
-- lens, observer -- source, and lens -- source, respectively,
and a concordance cosmological model is assumed with 
$\Omega_{\mathrm{tot}} = 1$, 
$\Omega_{\mathrm{matter}} = 0.3$, 
$\Omega_{\mathrm \Lambda} = 0.7$).

This length scale translates into an angular scale of:
\begin{equation}
	 \theta_E = {R_E \over D_S}  
		\approx  2.2 \times 10^{-6} \sqrt {M \over M_\odot} 
		~\rm  h_{75}^{-0.5} ~arcsec. 
\end{equation}
It is obvious that image splittings on such small 
angular scales can not be observed directly. 
What makes microlensing observable
in the first place is the fact that observer, 
lens(es) and source move relative to each
other. Due to this relative motion, the  microimage configuration 
changes with time, and so does the total magnification, i.e. the
combined fluxes of all the  microimages which make up 
the macro-image. This change
in magnification over time (lightcurve) can be measured:
microlensing is a dynamical phenomenon. 

There are two time scales involved: the standard lensing
time scale $t_E$ is the time it takes the lens (or the source)
to cross a length equivalent to the
Einstein radius, i.e.

\begin{equation}
	 t_E = (1+z_L) {R_E \over v_\perp}    
	\approx 25 \sqrt {M \over M_\odot}~v_{600}^{-1} ~\rm h_{75}^{-0.5} ~years, 
\end{equation}
where the same typical assumptions are made as above, 
and the relative transverse velocity\footnote{For simplicity
we assume here that all the transverse motion is done
by the lensing galaxy}  $v_{600}$ is parametrized in 
units of $v_\perp = 600$km/sec and for a Hubble constant 
$H_0 =  75 h_{75}$ km/sec/Mpc. 
This time scale results in relatively                large values. 
However, microlensing fluctuations are expected (and observed, see
Wo\'zniak et al. 2000a,b) on much shorter time scales.
The magnification distribution is highly non-linear with 
sharp caustic lines separating  regions of low and high magnification. 
Often, the density of caustics is quite high, so that a source
may encounter half a dozen or more caustic lines within the
length of one Einstein radius. 
When a source crosses a caustic, we can
expect a large change in magnification (and correspondingly in 
position) within the time it takes the source to
cross its own radius.
So the relevant time scale is the ``crossing time'':

\begin{eqnarray}
	 t_{cross} 
&=& (1+z_L)  {R_{source} \over v_\perp  (D_S/D_L)  } \nonumber \\
%&\approx&   0.58~R_{15} ~v_{600}^{-1} ~(D_S/ D_L)^{-1} ~\rm  h_{75}^{-0.5} ~years \nonumber \\
&\approx&   0.58~R_{15} ~v_{600}^{-1}~\rm  h_{75}^{-0.5} ~years \nonumber \\
&\approx&   210~R_{15} ~v_{600}^{-1} ~\rm  h_{75}^{-0.5} ~days.
\end{eqnarray}
Here the quasar size $R_{15}$ is parametrized in units of $10^{15}$cm.

\subsection{The special case of Q2237+0305}

The quadruple quasar Q2237+0305 
(Huchra et al. 1985; Irwin et al. 1989;
Wambsganss et al. 1990; Wyithe et al. 2000a,b; 
Wo\'zniak et al 2000a,b)
is a very special and favorable case and of particular
interest to microlensing studies.  It was the first system
in which microlensing was discovered (Irwin et al. 1989). Subsequently
it received a lot of attention, both 
observational
(Corrigan et al. 1991, Ostensen et al. 1996; 
Wo\'zniak et al 2000a,b) and 
theoretical
(Wambsganss et al. 1990; Wyithe et al. 2000a,b; Yonehara 2001).
Due to the fact that the lensing galaxy is so close 
($z_G = 0.039$, Huchra et al. 1985), 
the  physical and angular  Einstein 
radii  are considerably different from
the standard case treated above:

\begin{equation}
 r_{E, Q2237+0305}  \approx 
	1.6 \times 10^{16} \sqrt{M \over M_\odot} ~\rm h_{75}^{-0.5} ~cm, 
\end{equation}

\begin{equation}
	 \theta_{E, Q2237+0305} 
		\approx  7.3 \times 10^{-6} 
		\sqrt {M \over M_\odot} ~\rm  h_{75}^{-0.5} ~arcsec, 
\end{equation}

The resulting time scales (Einstein time and crossing time)
are much shorter than in almost all other
multiple quasars:  

\begin{equation}
	 t_{E, Q2237+0305} 
	\approx 8.7 \sqrt {M \over M_\odot} ~v_{600}^{-1}~\rm h_{75}^{-0.5} ~years, 
\end{equation}

\begin{eqnarray}
	 t_{cross, Q2237+0305}
%	&\approx&   0.05 ~R_{15} ~v_{600}^{-1} ~\rm   h_{75}^{-0.5} ~years \nonumber\\ 
	\approx   18 ~R_{15} ~v_{600}^{-1} ~\rm   h_{75}^{-0.5} ~days.
\end{eqnarray}
For that reason,
this quadruple system is ideally 
suited for microlensing studies. 
The length, time and angular scales for the ``typical case'' as
well as for the special case of Q2237+0305 are 
summarized in Table \ref{tab-cases}.

%%%%%%%%%%%%%%%%%%%%%%%%%%%%%%%%%%%%%%%%%%%%%%%%%%%%%%%%%%%%%%%%%%%
% Figure  1
%%%%%%%%%%%%%%%%%%%%%%%%%%%%%%%%%%%%%%%%%%%%%%%%%%%%%%%%%%%%%%%%%%%
%
	\begin{figure}
	\centering
	\includegraphics[width=87mm]{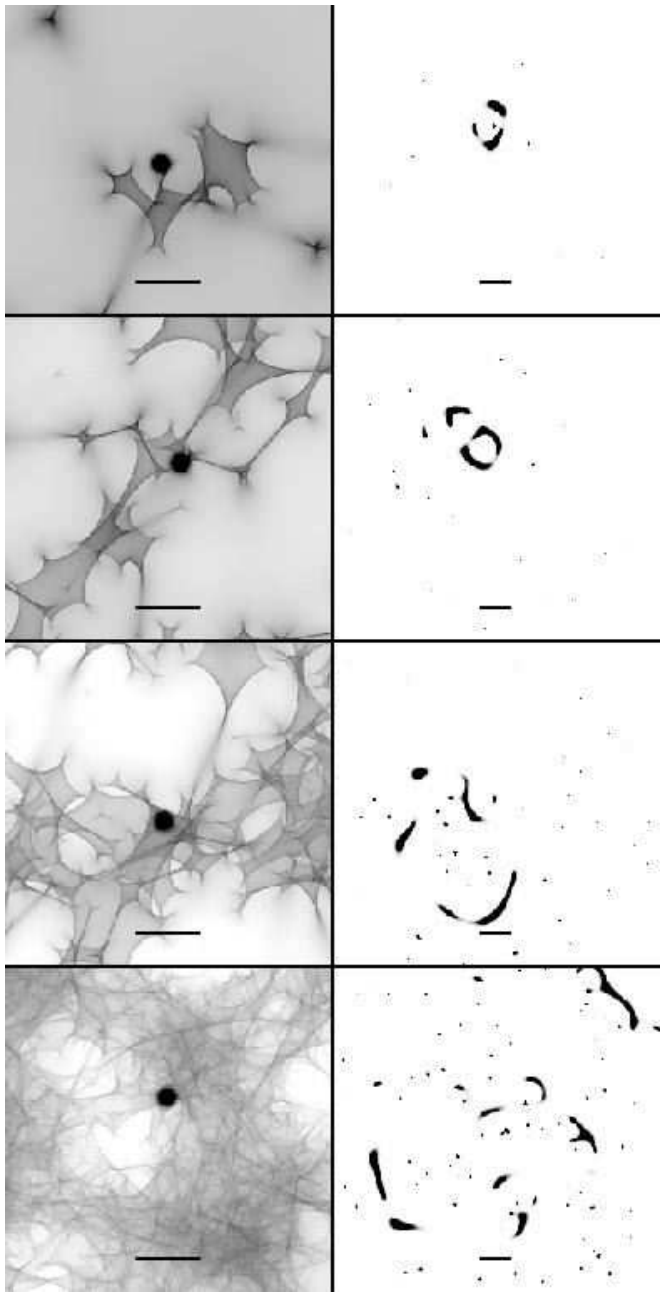}
	\caption{Illustration of source position in magnification 
	pattern/source plane (left) and 
	corresponding  microimage configuration 
	in the image plane, 
	for cases  $\kappa = 0.2$, $\kappa = 0.4$,
	$\kappa = 0.6$, and $\kappa = 0.8$ 
	(from top to bottom) and no external shear ($\gamma = 0.0$).
	The center-of-light position is marked with a plus sign.
	The horizonal bar indicates a length of two Einstein radii
	($2 \ R_E$).  
	Note the change in scale in the right hand panels.
	}
	\label{fig-microimages_1}
	\end{figure}

%%%%%%%%%%%%%%%%%%%%%%%%%%%%%%%%%%%%%%%%%%%%%%%%%%%%%%%%%%%%%%%%%%%
% Figure  2
%%%%%%%%%%%%%%%%%%%%%%%%%%%%%%%%%%%%%%%%%%%%%%%%%%%%%%%%%%%%%%%%%%%
%
	\begin{figure}
	\centering
	\includegraphics[width=87mm]{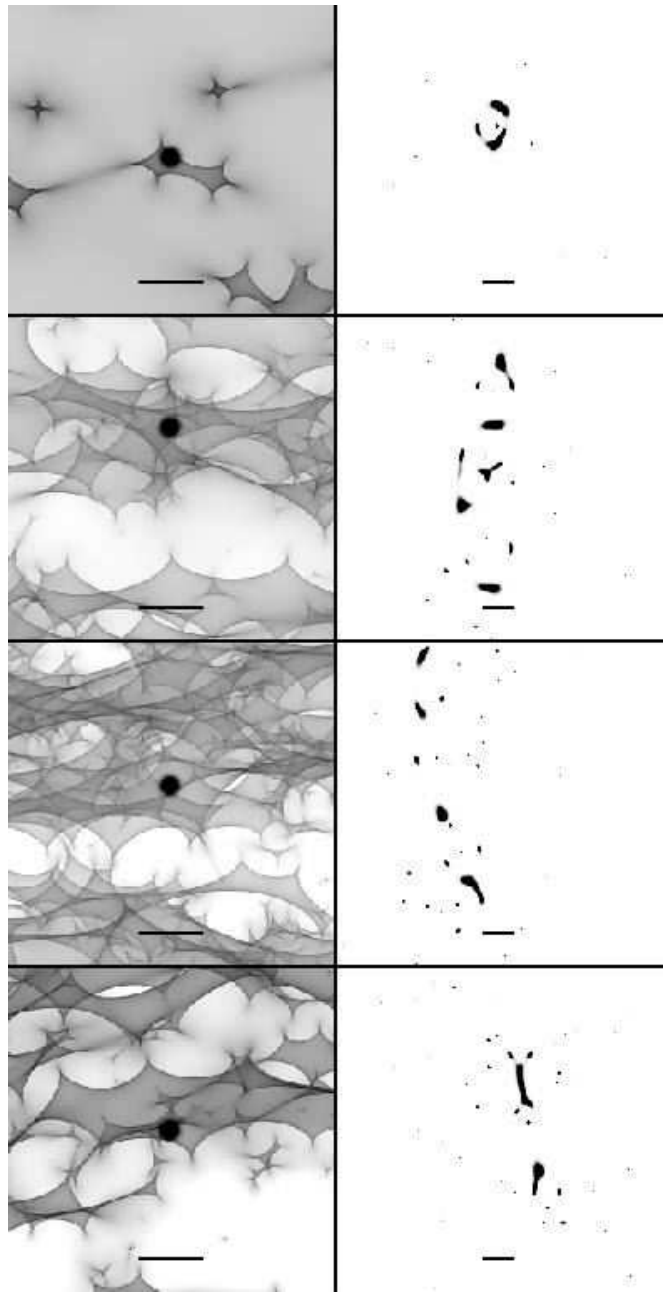}
	\caption{Same as Figure 1, here for the cases 
	{\it with}
	external shear  in the X-axis direction:
	$\kappa = \gamma = 0.2$, = 0.4,
	= 0.6, and = 0.8 (from top to bottom);
	the center-of-light position is marked with a plus sign,
	the horizonal bar indicates a length of two Einstein radii.
	}
	\label{fig-microimages_2}
	\end{figure}

\section{The Simulations}

\begin{figure}[t]
	\centering
	\includegraphics[width=87mm]{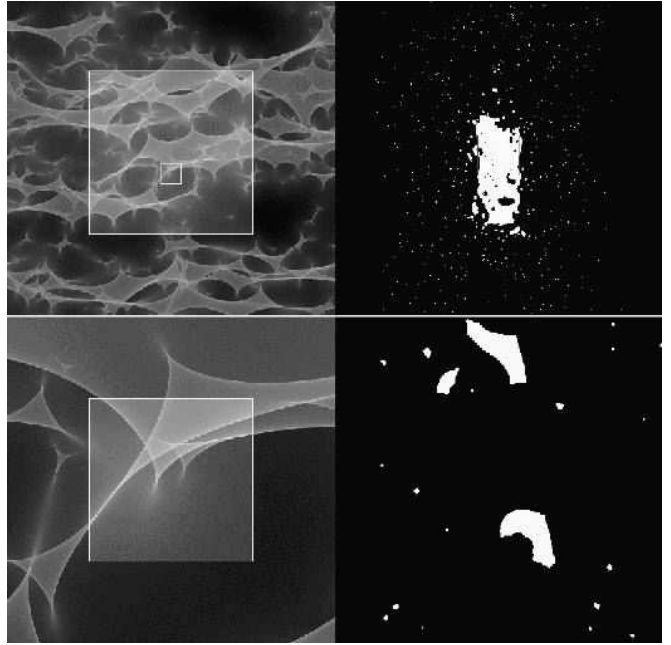}
	\caption{Illustration 
	of the ray shooting method, here for
	case $\kappa = 0.8$, $\gamma = 0.8$.
	Top left: 
	Magnification pattern in the source (quasar) plane
	with side length $L = 10 \ \theta_E$ (B configuration).
	%
	% !!! 
	%
%	{\bf 
	The large square indicates the region
	in the source plane for which all source positions
	were evaluated. 
	Top right: Corresponding microimage configuration of
        the large square ``source'' on the left,
	side length is 40 $\theta_E$: These are
	the parts of the sky (image plane) where
	light bundles originated from the square. Note the
	strong effect of shear in the Y-axis direction. 
	Bottom left: 
        The square here corresponds to the small square in the top 
        panel, zoomed in 8 times. 
	Bottom right: Corresponding micro-image configuration for
	the square on the left.
%	}
	%
	% !!!
	%
	}
	\label{fig-method}
	\end{figure}

In order to explore                      astrometric microlensing 
for a variety of realistic scenarios, 
we consider eight different cases, with the following
values of the dimensionless surface mass density:
$\kappa = 0.2, 0.4, 0.6$ and  0.8.
Each one of these we treated both without external shear 
($\gamma = 0$) and with external shear 
equal to the surface mass density
($\gamma = \kappa$), corresponding to an isothermal sphere model
for the lensing galaxy.
% 
% !!! 
% 
%{\bf 
The shear always acts along 
the X-axis in the simulations. This means that the caustics are expanded
in the X-direction and compressed in the Y-direction, as can be
seen for example in the top left panel of Fig.~\ref{fig-method}.
Horizontal bands appear on large scales (many Einstein
radii). The effect is not always obvious on small scales however
(sub-Einstein radii, bottom left panel of Fig.~\ref{fig-method}). 
%In contrast, for the micro-image configuration this means that 
Inversely, if we follow the light rays forward from source to 
observer (as opposed to backwards as in the simulations, which is
explained below), 
the micro-image configuration of a source will appear 
compressed in the X-direction %(not really obvious from Fig. 3) 
and expanded in the Y-direction (cf. the right hand
panels of Fig.~\ref{fig-method}).
%}
%
% !!!

For each of these eight cases, we produced a magnification pattern
in the source plane with a 
side length of $L_A = 20 \ R_E $, sampled on 
a $1000^2$ pixel  grid, i.e. $1 \  R_E$ is
covered by 50 pixels, or $1 \ {\mathrm{pix}} = 0.02 \ R_E$. 
This is our starting configuration A. 
The minimum source size we can consider is given by 
the pixel scale.
In order to explore smaller source
sizes as well (at the cost of smaller 
magnification patterns which may not be representative), 
we      zoom in  into the central part of the 
first configuration.   
We do higher resolution simulations (all on $1000^2$ pixel grids) 
in three steps of factors of two (configurations B, C and D,  respectively), 
resulting in magnification patterns with side lengths  of
$L_B = 10  R_E $, 
$L_C = 5 R_E $,  and 
$L_D = 2.5 R_E $,
respectively.
For the  highest resolution simulation
(configuration D), this means 
$R_E  = 400$ pixels, 
or 1 pix $= 0.0025 R_E$.
 
We used a modified version of the ray shooting 
code described in Wambsganss (1990, 1999).
In the original version, rays are followed backwards from
the observer through the lens plane (where all the deflections
are determined) to the source plane (where the light
rays are ``collected'' in pixels).
The density of rays then indicates the magnification
as a function of position in the source plane, often
displayed as color coded magnification patterns. 
Lightcurves can be obtained by 
convolving a given source profile with this two-dimensional
magnification map. However, all information about {\em where}
the rays originated from in the lens/image plane
is lost in this original algorithm. 
For the exploration of the centroid shift of the collection
of microimages, 
it is exactly this information that is required.
Hence we modified the code to record the positions of all  the
individual microimages brighter than a given magnification threshold.
To do this, we defined a second,
larger regular grid of $4000^2$ rays, which covers in the image plane
a region
of four times the angular side length of the magnification pattern
(i.e. $(80 \ \theta_E)^2$ for configuration A).
For a set of $ 1.6 \times 10^7$ test rays, 
we keep track of the positions in the
image plane as well as in the source plane. In this way, we
can find all the positions in the image plane that are
mapped onto a certain area in the source plane, and hence
identify all the microimages corresponding  to a particular
source position and size.

In order to get information on both 
the magnification and the 
 microimage locations of a finite source at a given
position, we convolved the
magnification pattern with a luminosity profile.
We used circular sources with Gaussian widths  
$\sigma =$ 2, 4, 8 and 16 pixels,
corresponding to physical sizes between
$0.04 \ R_E$ and $0.32 \ R_E$ in 
the starting configuration A, or
$1.5 \times  10^{15}$cm to  
$1.2 \times  10^{16}$cm in the typical case described in Section 2.1.
We did the same with the higher resolution magnification patterns,
configurations B, C and D.
 However, it turned out that the high resolution cases,  although well 
 suited for studying individual caustic crossings for
 small sources, are not quite large enough for statistical
 investigations.
For this reason, we
 restricted ourselves to cases A and B for the statistical 
 evaluations below.
The numerical values corresponding to the different
source sizes we used are tabulated in Table~\ref{tab-config}.

Using these simulations, we determined the 
positions of the individual microimages, 
   the center-of-light and the total
magnification 
of  the macro-image corresponding to a particular
source position and profile.
In Figs.~\ref{fig-microimages_1} and \ref{fig-microimages_2}, 
a  microimage situation is shown for each of the
eight cases considered: 
%
% !!! 
%
%{\bf 
$\kappa = 0.2, 0.4, 0.6$ and  0.8 with $\gamma = 0$ 
(Fig. \ref{fig-microimages_1}), and with $\gamma = \kappa$ 
acting along the X-axis direction (Fig. \ref{fig-microimages_2}).
%}
%
% !!!
%
In the left columns, a small part of the magnification
pattern is shown, with the source position and profile
superimposed. The panels on the right hand side show
the particular  microimage configurations, with the
light centroid indicated by a plus sign 
(cf. Paczy\'nski 1986).

%
%%%%%%%%%%%%%%%%%%%%%%%%%%%%%%%%%%%%%%%%%%%%%%%%%%%%%%%%%%%%%%%%%%%
% Table 2:
%%%%%%%%%%%%%%%%%%%%%%%%%%%%%%%%%%%%%%%%%%%%%%%%%%%%%%%%%%%%%%%%%%%
%
%__________________________________________________ One column table
%
%
   \begin{table}[t]
      \caption[]{Source sizes (Gaussian widths) used in
	our microlensing simulations (configurations A and B) 
        in units of pixels, 
	Einstein radii and in
      physical units for the typical lensed quasar and for the
      special case of Q2237+0305}
        \label{tab-config}
	\begin{tabular}{lcccc}
	\hline
	\noalign{\smallskip}
&      $\sigma$   & in $R_E$ & in cm          & in cm        \\
&      in pix  &          & (typical case) & (Q2237+0305) \\
	\noalign{\smallskip}
	\hline
	\noalign{\smallskip}
A                  &  16& 0.32 & $1.2 \times 10^{16}$ & $4 \times 10^{16}$\\
		   &  8 & 0.16 & $6.0 \times 10^{15}$ & $2 \times 10^{16}$ \\
                   &  4 & 0.08 & $3.0 \times 10^{15}$ & $1 \times 10^{16}$ \\
                   &  2 & 0.04 & $1.5 \times 10^{15}$ & $5 \times 10^{15}$ \\
	\noalign{\smallskip}
	\hline
	\noalign{\smallskip}
B                  & 4 & 0.04 & $1.5\times 10^{15}$ & $5  \times 10^{15}$ \\
		   & 2 & 0.02 & $7.5\times 10^{14}$ & $2.5\times 10^{15}$ \\
	\noalign{\smallskip}
	\hline
	\end{tabular}
\end{table}

%%%%%%%%%%%%%%%%%%%%%%%%%%%%%%%%%%%%%%%%%%%%%%%%%%%%%%%%%%%%%%%%%%%
% Figure 3
%%%%%%%%%%%%%%%%%%%%%%%%%%%%%%%%%%%%%%%%%%%%%%%%%%%%%%%%%%%%%%%%%%%
%

In Figure \ref{fig-method}, the method is  illustrated for
$\kappa = 0.8$ and $\gamma = 0.8$. 
The top left panel shows the
magnification pattern in the source (quasar) plane
with side length $L = 10 \ \theta_E$ (B configuration).
The grey scale indicates regions of different magnifications
in the source plane: the lighter the grey, 
the higher magnification.
The large square indicates the region in the source plane 
for which all source positions were evaluated.  
The top right panel shows the corresponding microimage 
configuration in the image plane: the
white regions are the parts of the image plane
where light bundles appear which originated from within the
square on the left (side length is 40 $\theta_E$). 
%{\bf 
In other words, it shows what a large square shaped source 
would look like to the observer.
The bottom panels show the same for
the smaller square region indicated 
in the top left panel, zoomed eight times.
The many isolated light patches indicate that microimages are
spread over a very large area in the image plane.
%}

\begin{figure}[t]
   \centering
   \includegraphics[width=87mm]{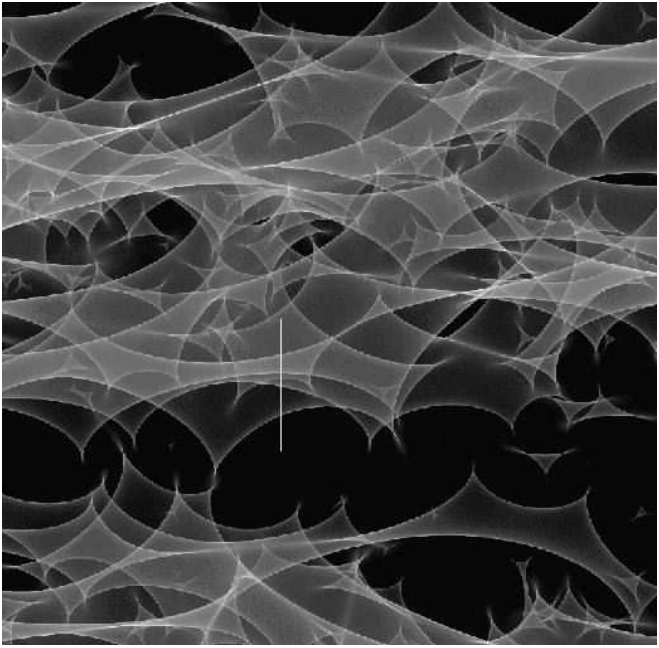}
   \caption{Illustration of a quasar microlensing scenario
	(here for lensing parameters $\kappa = 0.6$ and $\gamma = 0.6$,
	sidelength about 10 $R_E$):
	The gray scale indicates regions of different magnifications
	in the source plane.
	The straight vertical line marks the track of the 
	quasar relative to the magnification pattern.
	The total length of the track is 2.0 $R_E$.
	}
	\label{fig-magpat_light}
   \end{figure}

\begin{figure}[t]
   \centering
   \includegraphics[width=87mm]{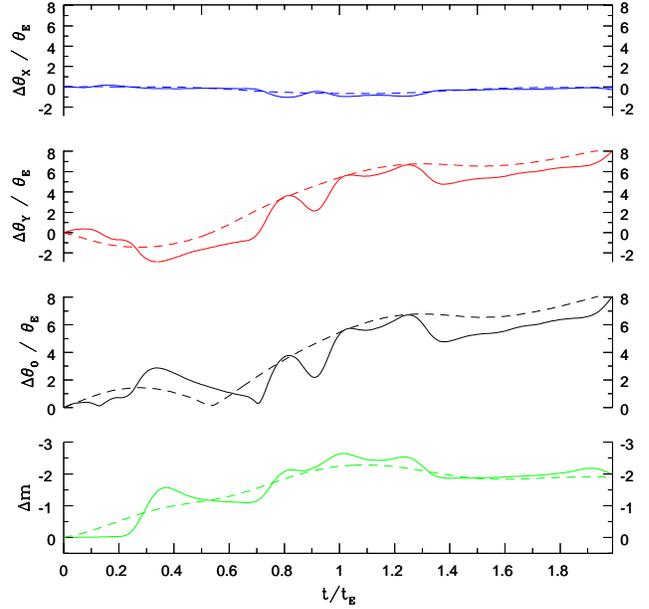}
   \caption{The four lines indicate from top to bottom:
a) 	$\Delta \theta_X$,
	the X-coordinate 
	of the quasar relative to the starting position
	as a function of time;
b) same for the Y-coordinate;
c) the absolute value of the positional shift
	$|\Delta \theta|$
	relative to the starting position as a function of time;
and d) the corresponding light curve of the quasar
for the example track shown in Figures \ref{fig-magpat_light}.
The solid lines correspond
to a source size $\sigma = 0.02 R_E$ 
(typically about $6.4 \times 10^{14} \sqrt{M/M_\odot} h_{75}^{-0.5}$cm)
the dashed lines to $\sigma= 0.16 R_E$ 
(typically about $5.1 \times 10^{15} \sqrt{M/M_\odot} h_{75}^{-0.5}$cm).
}
	\label{fig-lightcurve}
   \end{figure}

Figure \ref{fig-magpat_light} shows a quasar microlensing scenario
with microlensing parameters $\kappa = 0.6$ and $\gamma = 0.6$,
and side length 10 $\theta_E$ (B configuration).
The straight vertical white line marks the track of the 
quasar motion relative to the magnification pattern; the length of 
the path is 2.0 $R_E$. 
For this particular track (followed from the lowest part upwards), 
Figure \ref{fig-lightcurve} shows from top to bottom:
the X- and Y-coordinates
	of the quasar relative to the starting position
	($\Delta \theta_X$ and $\Delta \theta_Y$) as a function of time;
the absolute value of the positional shift
	($|\Delta \theta|$)
	relative to the starting position as a function of time;
and the corresponding light curve ($\Delta m$) of the quasar.
The solid and dotted lines correspond
to  two different values of the source size:
$\sigma = 0.04 R_E$ (4 pixels in B configuration) and
$\sigma= 0.16 R_E$ (16 pixels in B configuration), respectively.
The track  in Fig. \ref{fig-magpat_light} 
starts in a region of low
magnification which is taken as the zero point of the magnitude
scale on the lowest panel in \ref{fig-lightcurve}.

\begin{figure}
   \centering
   \includegraphics[width=87mm]{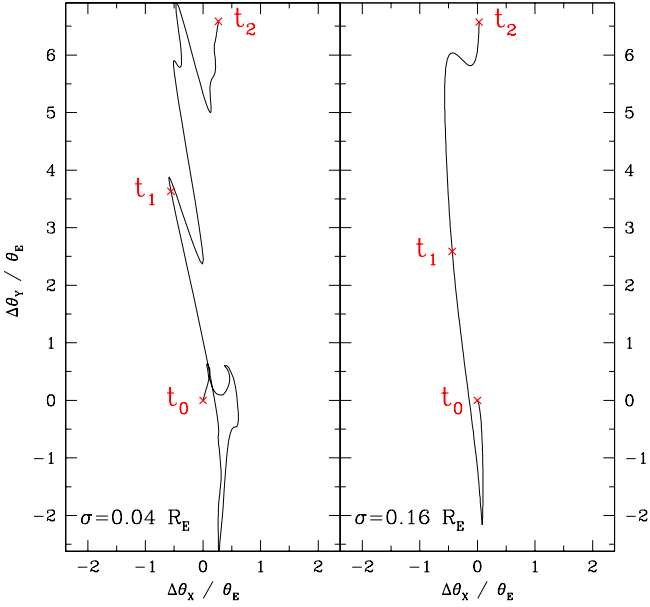}
   \caption{Effect of the source size on the centroid shift
   for the example track shown in Fig.~\ref{fig-magpat_light}
   and \ref{fig-lightcurve} ($\kappa = 0.6$ and $\gamma = 0.6$).
   The two panels show the astrometric tracks that would
   occur for the linear quasar motion shown in Fig.~\ref{fig-magpat_light} 
   for two different source sizes: 
   $\sigma=$ 0.04 and 0.16 $R_E$ 
   (4 and 16 pixels in configuration B, respectively). 
   The total length of the track is 
   2.0 $t_E$ or about 15 years for the quasar Q2237+0305.
   The  labels $t_0$, $t_1$ and $t_2$ correspond 
   to the starting, central and final positions, respectively.
	}
         \label{fig-centroid_shift}
    \end{figure}

\begin{figure}
   \centering
   \includegraphics[width=87mm]{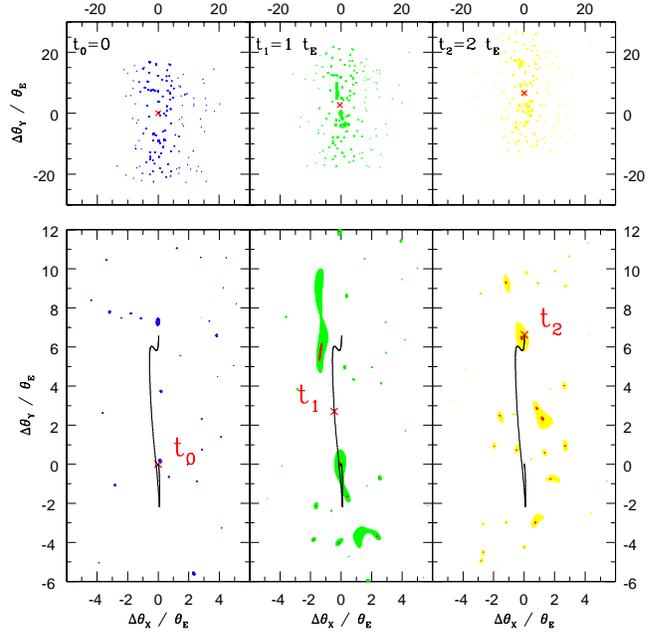}
   \caption{Top: Three snapshots of the ensemble of microimages
   assuming $\sigma= 0.16 R_E$. 
   %
   %!!!
   %
%   {\bf 
   All bright microimages are plotted 
   up to a radius of 3 $\sigma$ (due to the finite resolution
   of the plotting routine, some of the faint microimages
   are represented larger than they actually are. 
%   }
   %
   %!!! 
   %
   Left to right: starting, middle et final positions of the track 
   in Fig.~\ref{fig-magpat_light}, corresponding to t=0.0, 1.0 
	and 2.0 $t_E$, respectively
  (note the angular scale: the microimage distribution 
   covers about 30 $\theta_E \times \ 40 \ \theta_E$!).
   Bottom: zoomed central regions;
	the respective positions of the center-of-light for 
	the three
	epochs are marked as crosses on the track of
	the center-of-light
   	(with the very central  source parts corresponding to 
   	0.5 $\sigma$ indicated in red).
	}
         \label{fig-snapshots}
    \end{figure}

The two panels in Figure \ref{fig-centroid_shift} represent
the centroid shift for the two different source sizes.
The whole track has a length of 
2.0 $t_E$ ($\sim$ 50 and 15 years for the 'typical' lensing case
and the Q2237+0305 case, respectively). The labels
$t_0$, $t_1$ and $t_2$ correspond 
to the starting, middle and final positions, respectively.
Snapshots of the ensemble of microimages at times 
$t_0$, $t_1$ and $t_2$ 
are shown in Figure \ref{fig-snapshots} assuming $\sigma= 0.16 R_E$.
%
% !!! 
%
%{\bf
All microimages are plotted up to a radius of 3 $\sigma$,
without accounting for the declining source profile.
On average, the brightness of the macroimage declines as
the fourth power of distance to the center of light 
(Katz, Balbus, Paczy\'nski 1986). 
On our plots, due to the finite resolution, some of the
distant and faint microimages appear bigger  than
they are. 
This figure simply aims at illustrating the very 
large spatial spread of the microimages,
covering roughly $30 \ \theta_E \times 40 \ \theta_E$. 
%}
%
% !!!
%
The lower panels show enlargements of the central parts with the
path of the center-of-light superimposed.

%
%%%%%%%%%%%%%%%%%%%%%%%%%%%%%%%%%%%%%%%%%%%%%%%%%%%%%%%%%%%%%%%%%%%
% Figure 4
%%%%%%%%%%%%%%%%%%%%%%%%%%%%%%%%%%%%%%%%%%%%%%%%%%%%%%%%%%%%%%%%%%%
%

%
%%%%%%%%%%%%%%%%%%%%%%%%%%%%%%%%%%%%%%%%%%%%%%%%%%%%%%%%%%%%%%%%%%%
% Figure 5
%%%%%%%%%%%%%%%%%%%%%%%%%%%%%%%%%%%%%%%%%%%%%%%%%%%%%%%%%%%%%%%%%%%
%

%
%%%%%%%%%%%%%%%%%%%%%%%%%%%%%%%%%%%%%%%%%%%%%%%%%%%%%%%%%%%%%%%%%%%
% Figure 6
%%%%%%%%%%%%%%%%%%%%%%%%%%%%%%%%%%%%%%%%%%%%%%%%%%%%%%%%%%%%%%%%%%%
%

%
%%%%%%%%%%%%%%%%%%%%%%%%%%%%%%%%%%%%%%%%%%%%%%%%%%%%%%%%%%%%%%%%%%%
% Figure 7
%%%%%%%%%%%%%%%%%%%%%%%%%%%%%%%%%%%%%%%%%%%%%%%%%%%%%%%%%%%%%%%%%%%
%

\section{Results and Discussion}

\subsection{Microlens-induced positional offsets and corresponding
magnitude changes}

%
%%%%%%%%%%%%%%%%%%%%%%%%%%%%%%%%%%%%%%%%%%%%%%%%%%%%%%%%%%%%%%%%%%%
% Figure 8
%%%%%%%%%%%%%%%%%%%%%%%%%%%%%%%%%%%%%%%%%%%%%%%%%%%%%%%%%%%%%%%%%%%
%

  \begin{figure}
  \centering
   \includegraphics[width=87mm]{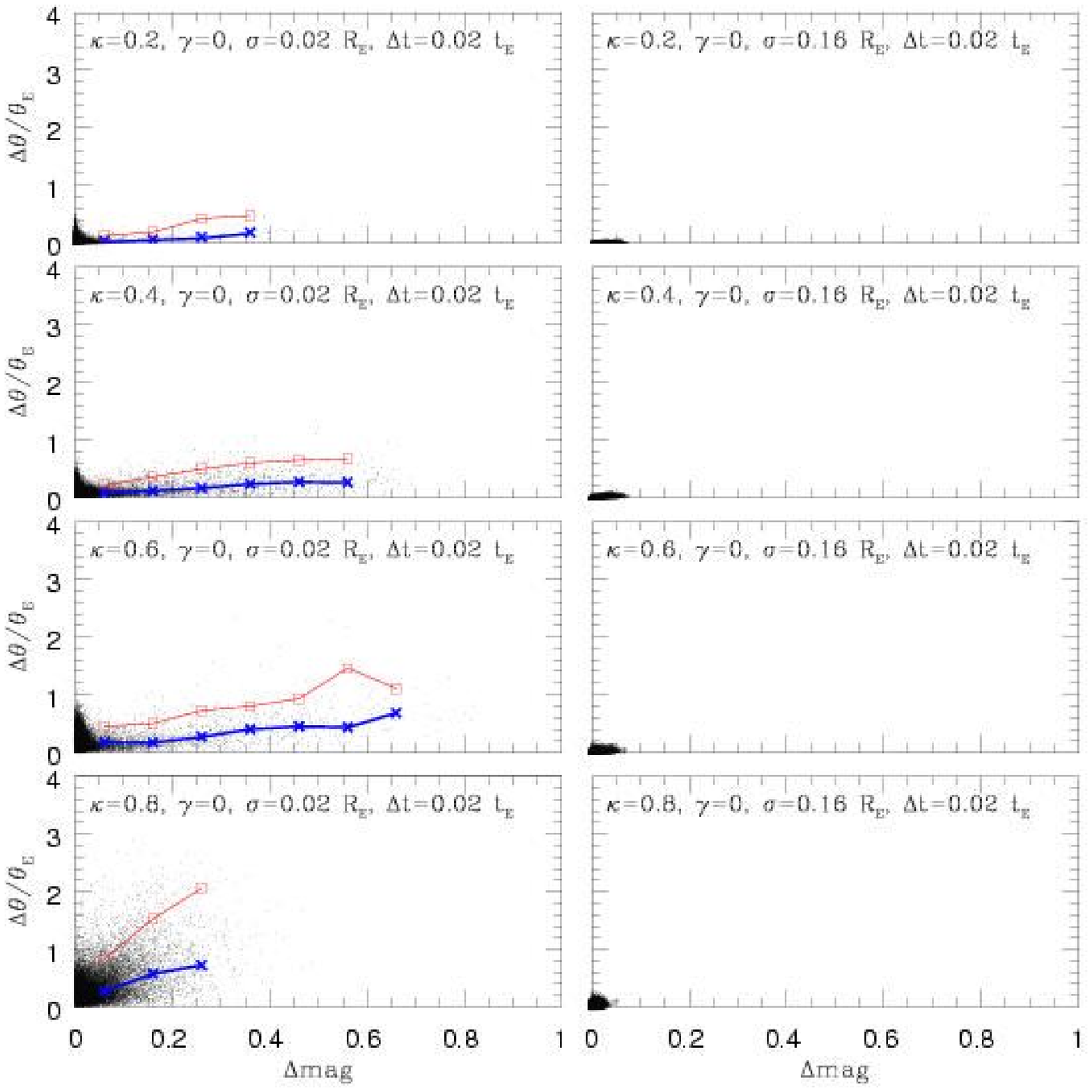}
   \caption{Offset in position $|\Delta \theta|$ versus
   offset in apparent brightness $|\Delta m|$,
   for two simulated measurements separated
        by $\Delta t = 0.02 t_E$.
	This time interval corresponds to 
	about half a year for the ``typical'' lensing case and
	to two months  for the case of Q2237+0305 (cf. Table \ref{tab-cases}).
        From top to bottom, the panels represent
	 the cases
        $\kappa=$ 0.2, 0.4, 0.6 and 0.8, respectively, without
	external shear ($\gamma=0.0$).
	The source sizes
        are $\sigma = 0.02 R_E$ (left column),
        and $0.16 R_E$ (right column).
        The median of the $|\Delta \theta|$ distribution as
        a function of $|\Delta m|$ is shown as  a thick
	line (points marked as crosses), 
	the 95th-percentile is indicated as a thin line 
	(and squares).
        }
         \label{fig-mag_offset1_t0.02}
  \end{figure}

%%%%%%%%%%%%%%%%%%%%%%%%%%%%%%%%%%%%%%%%%%%%%%%%%%%%%%%%%%%%%%%%%%%
% Figure 9
%%%%%%%%%%%%%%%%%%%%%%%%%%%%%%%%%%%%%%%%%%%%%%%%%%%%%%%%%%%%%%%%%%%

  \begin{figure}
  \centering
   \includegraphics[width=87mm]{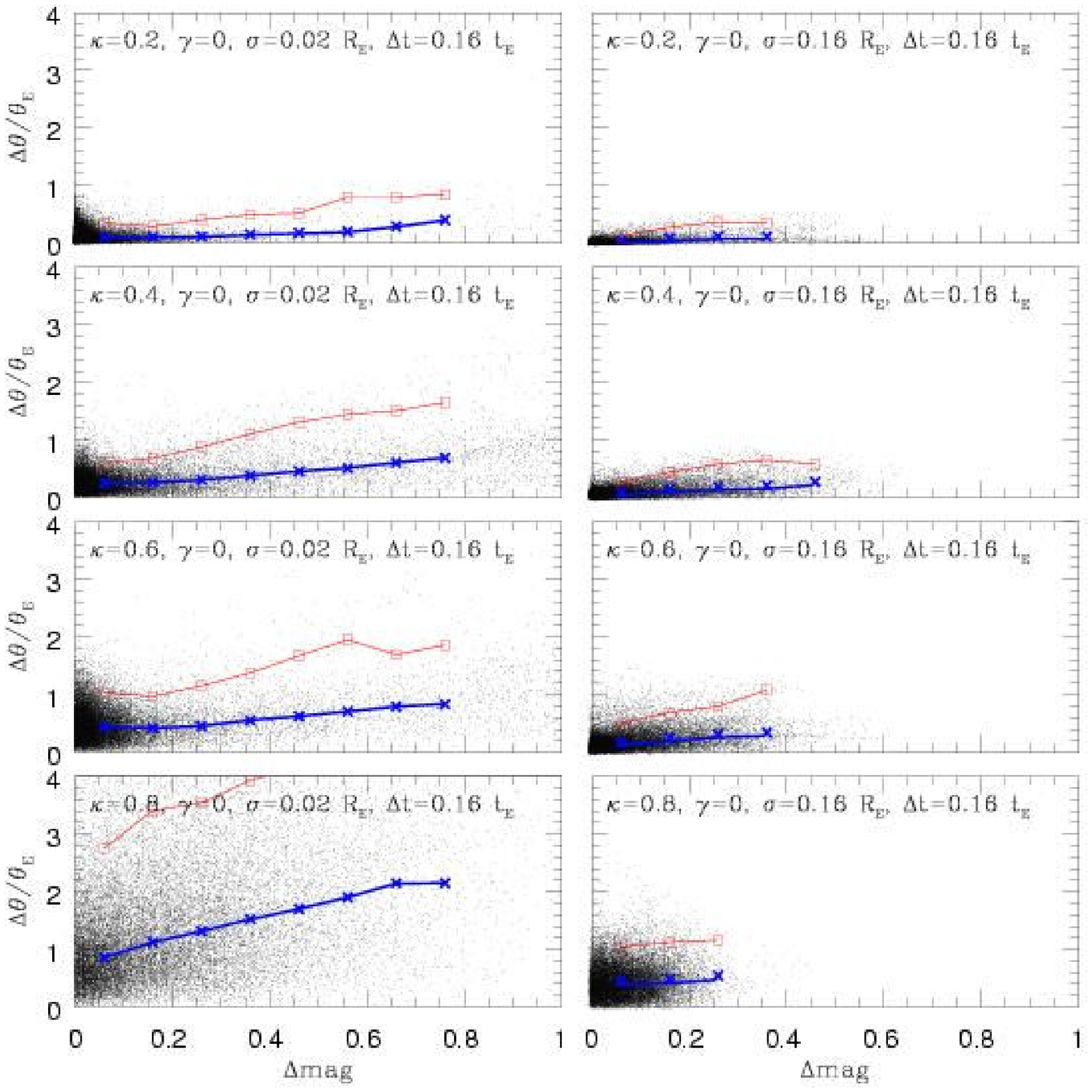}
   \caption{Same as Fig.~\ref{fig-mag_offset1_t0.02}, but here for
   two measurements separated
        by a larger time interval $\Delta t = 0.16 t_E$
 	(corresponding to  about four years in the ``typical'' 
	case, and 1.5 years for the case of Q2237+0305, cf. Table \ref{tab-cases}):
	source sizes are $\sigma = 0.02 R_E$ (left column),
        and $0.16 R_E$ (right column);
        median of $|\Delta \theta|$ as
        function of $|\Delta m|$ is thick
	line (points are crosses), 
	95th-percentile is thin line (squares).
        }
         \label{fig-mag_offset1_t0.16}
  \end{figure}

\begin{figure}
%   \centering
   \includegraphics[width=87mm]{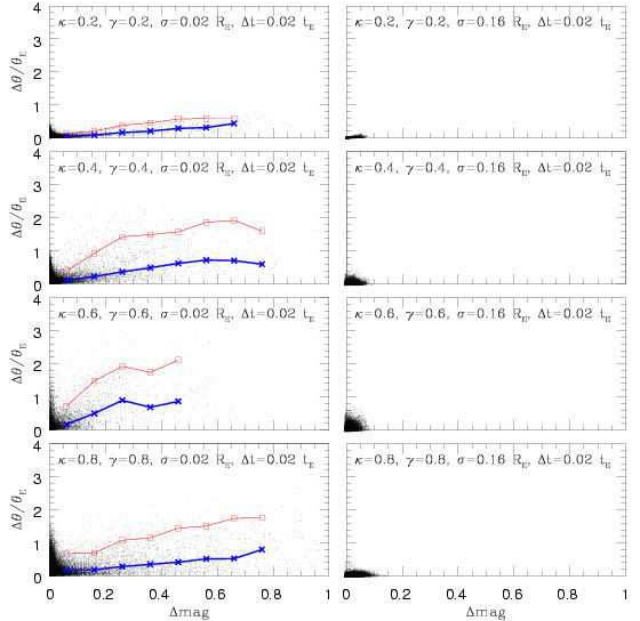}
   \caption{Same as Fig.~\ref{fig-mag_offset1_t0.02}, but here for the 
	cases  with external shear 	 
	$\gamma=\kappa = 0.2, 0.4. 0.6, 0.8 $ from top to bottom:
	source sizes are $\sigma = 0.02 R_E$ (left column),
        and $0.16 R_E$ (right column);
        median of $|\Delta \theta|$ as
        function of $|\Delta m|$ is thick
	line (points are crosses), 
	95th-percentile is thin line (squares).
	}
         \label{fig-mag_offset2_t0.02}
   \end{figure}

\begin{figure}
   \centering
   \includegraphics[width=87mm]{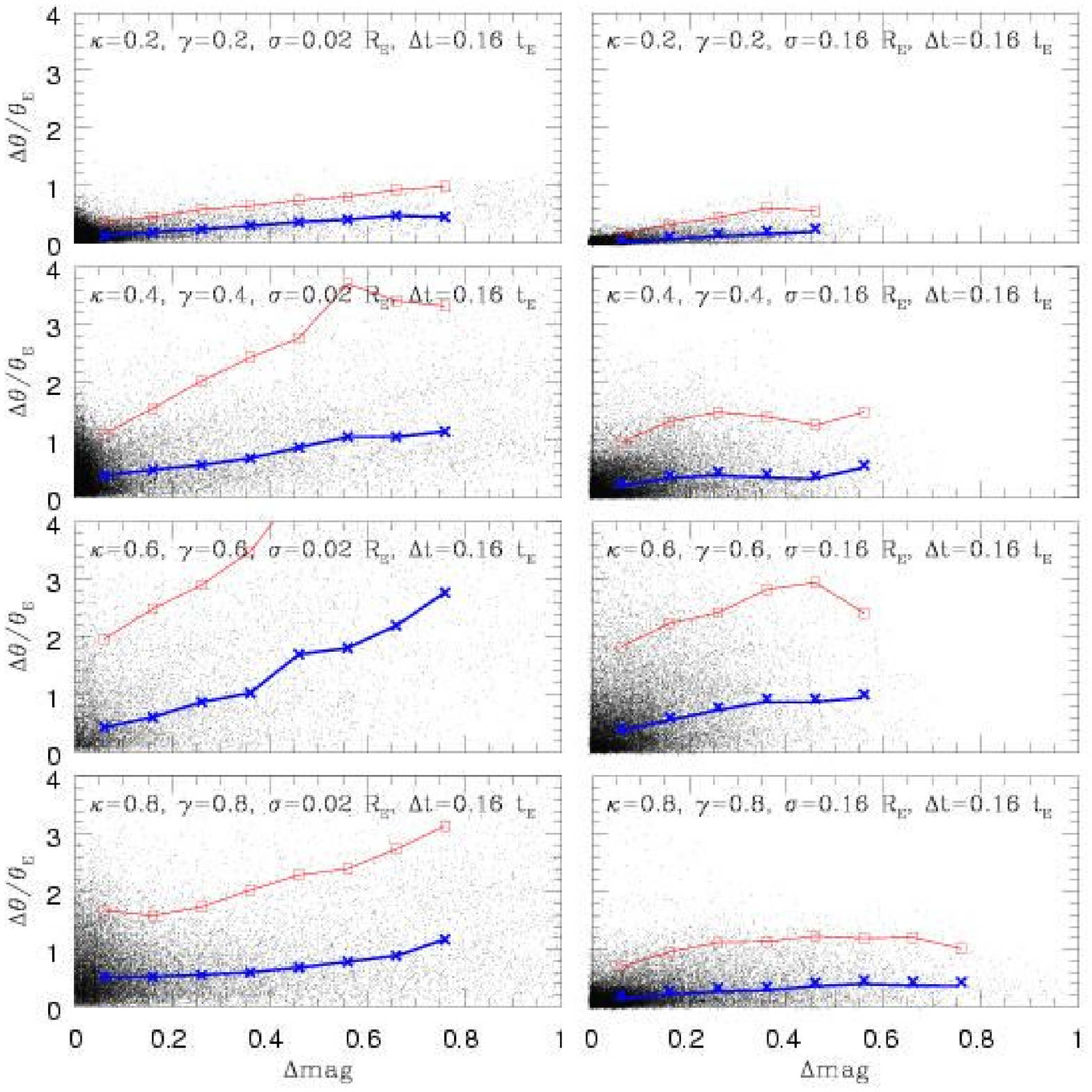}
   \caption{Same as Fig.~\ref{fig-mag_offset2_t0.02}, but here for the
   two measurements separated
        by a larger time interval $\Delta t = 0.16 t_E$:
	source sizes are $\sigma = 0.02 R_E$ (left column),
        and $0.16 R_E$ (right column);
        median of $|\Delta \theta|$ as
        function of $|\Delta m|$ is thick
	line (points are crosses), 
	95th-percentile is thin line (squares).
}
         \label{fig-mag_offset2_t0.16}
   \end{figure}

In order to evaluate the correlations between 
the magnitude changes and the positional changes of a 
microlensed quasar,
we simulated pairs of measurements separated by  
time intervals of 
0.02 $t_E$ (Figures \ref{fig-mag_offset1_t0.02},
	\ref{fig-mag_offset2_t0.02}) and 
0.16 $t_E$ (Figures \ref{fig-mag_offset1_t0.16},
	\ref{fig-mag_offset2_t0.16}), respectively.
These values correspond
to  about half a year and four years in the ``typical'' case, and
to two months  and 1.5 years in the case of Q2237+0305 (cf. Table \ref{tab-cases}).
We placed the source at a random position in the magnification pattern
and chose a second position at a distance
of $0.02 R_E$ or $0.16R_E$ either parallel or perpendicular to
the action of the external shear                    (X-axis).
We determined differences in the magnifications $|\Delta m|$
and the center-of-light positions $|\Delta \theta|$ between these
two source positions. Each pair of measurements 
($|\Delta m|$, $|\Delta \theta|$) is represented
as a point in the various panels of 
Figures  \ref{fig-mag_offset1_t0.02} to \ref{fig-mag_offset2_t0.16}.
In each panel, the offset in position $|\Delta \theta|$ 
is shown against the corresponding offset in magnitude $|\Delta m|$  
for about 20,000  such measurement pairs.
In the left columns of all four figures, 
a small Gaussian source size of 
	$\sigma = 0.02 R_E$ is assumed, in the right column the
	source size  is $\sigma = 0.16 R_E$.
In each panel, the        thick (lower) line indicates the median of 
$|\Delta \theta|$ as a function of $|\Delta m|$, and the thin (upper)
line       shows the 95th-percentile: 5\% of all simulations 
for a given $|\Delta m|$ would result in offsets $|\Delta \theta|$ 
which are above this thin line.

In Figure \ref{fig-mag_offset1_t0.02}, the four lensing situations
without external shear are considered: $\kappa = 0.2, 0.4, 0.6, 0.8$ 
(from top to bottom) with a small time step    $\Delta t = 0.02 t_E$.
In all panels, the majority  of the points cluster near the
origin. 
This is easily understandable: 
the position of the source relative to the caustics has
not changed by much during the 
short time interval, so the changes both in magnification
and in the centroid position tend to be  small.
For                 small sources (left panels), however,       
$|\Delta m|$ and $|\Delta \theta|$ can reach relatively 
large values with 
the median lines indicating an almost linear statistical
relation.
The slope of these median lines
slightly increases with increasing surface mass density $\kappa$.
The time step 
 corresponds roughly to the crossing time:
$\Delta t = 0.02 t_E \approx t_{\mathrm{cross}}$, 
and there are indeed
cases with easily measurable magnitude fluctuations
($|\Delta m| > 0.2$ mag), which result in center-of-light offsets of
about 0.5 $\theta_E$.
In the right panel, there are no significant changes in either
magnification or center-of-light position, because the
source has moved only a fraction of its own diameter
($\Delta t < t_{\mathrm{cross}}$).

In Figure \ref{fig-mag_offset1_t0.16}, the same is shown for a larger
time step $\Delta t = 0.16 t_E$.                  Many more
points are now spread towards larger offsets and
larger magnification changes. 
For the small source, median 
values of $|\Delta \theta| \ge 1 \theta_E$ are reached
in  the highest surface
mass density cases. In fact, the 95th-percentile line for
the $\kappa = 0.8$ case indicates that for magnitude
changes    $|\Delta m | > 0.4$mag, 5\% of the all cases
result in center-of-light offsets larger  than 4$\theta_E$. 
For the large source (right column) the expected 
offsets are still quite moderate, with median values  of
about 0.5 $\theta_E$. 
This is not too surprising, because the time interval 
corresponds to just about the crossing time for the 
source.

%
%%%%%%%%%%%%%%%%%%%%%%%%%%%%%%%%%%%%%%%%%%%%%%%%%%%%%%%%%%%%%%%%%%%
% Figure 10
%%%%%%%%%%%%%%%%%%%%%%%%%%%%%%%%%%%%%%%%%%%%%%%%%%%%%%%%%%%%%%%%%%%
%

%
%%%%%%%%%%%%%%%%%%%%%%%%%%%%%%%%%%%%%%%%%%%%%%%%%%%%%%%%%%%%%%%%%%%
% Figure 11
%%%%%%%%%%%%%%%%%%%%%%%%%%%%%%%%%%%%%%%%%%%%%%%%%%%%%%%%%%%%%%%%%%%
%

Figures \ref{fig-mag_offset2_t0.02} and \ref{fig-mag_offset2_t0.16} 
 contain the same diagrams 
for the cases with external shear, $\kappa = \gamma = 
0.2, 0.4, 0.6, 0.8$ (from top to bottom).
Larger offsets are reached here than in the corresponding
scenarios without shear. 
Especially the two middle rows with   $\kappa = \gamma = 0.4$
and 0.6  --  which best represent the typical values of
convergence and shear of a multiply imaged quasar -- produce
values of positional offset a factor of two higher than
the corresponding cases without shear: 
For short time steps and small sources, 
medians of 
$|\Delta \theta| \approx 1 \theta_E$ and 95th-percentiles
of $|\Delta \theta| \approx  2 \theta_E$ (small source, left
column) are reached.
For smaller values of the surface mass density, the
caustic density and hence the number of microimages
is not large enough to produce big positional offsets, whereas
for higher values of   $\kappa$, the density of
caustics is so high that an additional microimage pair  only
produces a  small fluctuation.

Results are even more dramatic when assuming 
a larger time step ($\Delta t = 0.16 t_E$), as displayed
in Figure  \ref{fig-mag_offset2_t0.16}:  
median values of $|\Delta \theta|$ between 
0.5$ \theta_E$ and
3$ \theta_E$ for the small source size (left column), and
significant median values of 
$|\Delta \theta| \approx 0.5\theta_E$ to $1\theta_E$
even for the large source (right column).

\subsection{Microlens-induced positional offset as a function
of time (with  magnitude thresholds)}

\begin{figure}
   \centering
	\includegraphics[width=87mm]{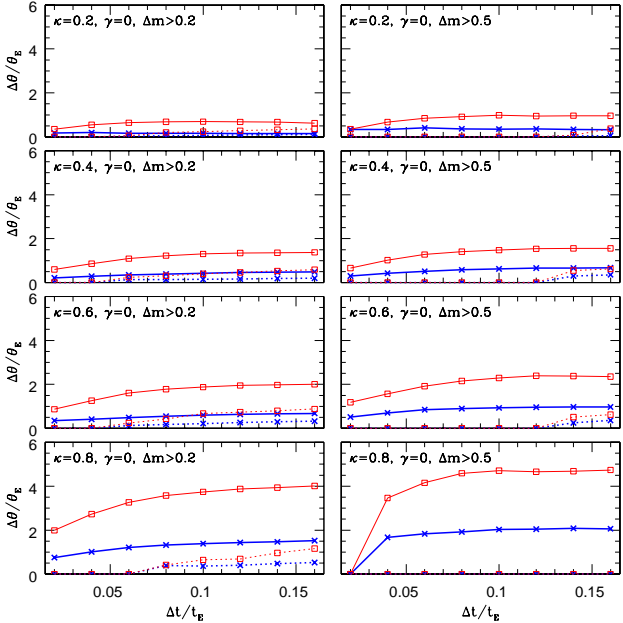}
	\caption{The median (thick lines with cross symbols) and 
	95th percentile (thin lines with square symbols) 
	of the $|\Delta \theta|$ distribution as a function of 
	$\Delta t$ for
    	$|\Delta m| > 0.2$ (left panels) and 
	$|\Delta m| > 0.5$ (right panels). 
        The solid and dotted lines represent the two source sizes
   	$\sigma = 0.02 R_E$ and $0.16 R_E$, respectively.
	From top to bottom, the panels correspond to 
	increasing surface mass density: 
	$\kappa=$ 0.2, 0.4, 0.6 and 0.8, respectively, with 
	no external shear ($\gamma=0$).
	}
         \label{fig-time_offset1}
   \end{figure}

\begin{figure}
   \centering
	\includegraphics[width=87mm]{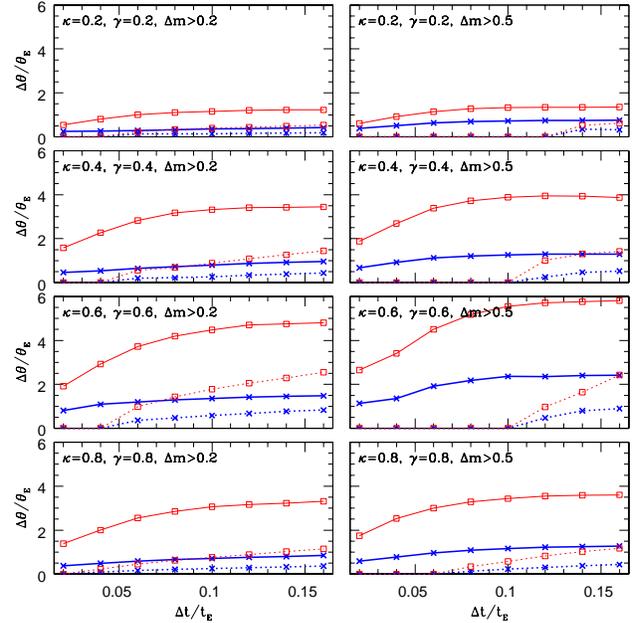}
	\caption{Same as Fig.~\ref{fig-time_offset1}, here for the cases
	with external shear $\gamma=\kappa = 0.2, 0.4, 0.6, 0.8$ (from
	top to bottom):
	median (thick lines, crosses) and 
        95th-percentile (thin lines,  squares) 
        of $|\Delta \theta|$ distribution as a function of 
        $\Delta t$ for
        $|\Delta m| > 0.2$ (left panels) and 
        $|\Delta m| > 0.5$ (right panels);
        solid lines for small source ($\sigma = 0.02 R_E$),
	dotted lines  for large source ($0.16 R_E$), respectively.
	}
         \label{fig-time_offset2}
   \end{figure}

We evaluated the shifts in center-of-light positions for 
time intervals intermediate between 
$0.02 \le \Delta t/t_E \le 0.16$ assuming a threshold for
the magnitude fluctuations.
In Figures  \ref{fig-time_offset1} and \ref{fig-time_offset2}, the
median and 95th-percentile offsets are shown as
a function of increasing time step $\Delta t$,
for 
$|\Delta m|\ge 0.2$ mag  (left panels) and
$|\Delta m|\ge 0.5$ mag  (right panels).  
This was done with the following idea in mind: since
very accurate measurements of quasar image positions are
``expensive", it is unlikely that all lensed quasar candidates
can be astrometrically monitored. However, photometric monitoring
is comparably cheaper. 
So ideally, one could determine the positions of
the most promising multiple quasars with high accuracy once,
then monitor them photometrically, and whenever a large
microlens-induced magnitude change has been detected,
a second astrometric measurement should be performed.

In each panel of Figure  \ref{fig-time_offset1}, the two sets of curves
show the median (thick lines) and the 95th-percentile (thin lines)
of the $|\Delta \theta|$ distribution 
as a function of  $\Delta t$  for
the small source ($\sigma = 0.02 R_E$, solid) and
the large source ($\sigma = 0.16 R_E$, dotted). All curves show
basically the same behaviour: the $|\Delta \theta|$
values first increase  with increasing $\Delta t$, then
     flatten out. This behaviour shows that many or most
jumps in magnitude and position are dominated by 
one fold-caustic crossing. 
The
time scale is dominated by $t_{\mathrm{cross}}$. For a slightly
larger time interval, the offset does not increase significantly 
any more. 
Only for much larger $\Delta t$'s 
allowing for additional caustic crossings, 
would another increase in $|\Delta \theta|$ 
be expected.  Offsets of more than
100 microarcseconds could indeed be reached this way,
but the characteristic time scale would be depressingly
large (many decades!).

The qualitative behaviour of the cases with external
shear displayed in Figure  \ref{fig-time_offset2} is similar to
those without shear. 
However, the expected offset values are significantly higher 
here, as already seen in Figs. 
\ref{fig-mag_offset2_t0.02} and
\ref{fig-mag_offset2_t0.16}:
again, the cases $\kappa = \gamma = 0.4$ and 0.6 appear most promising
(second/third row), with median values of between  1   $\theta_E$
and 2 $\theta_E$,
and
95th-percentiles of 4 $\theta_E$ or higher. 
These     values translate into about 7 to 15 microarcseconds  (median)
and $>$ 30 microarcseconds (95th percentile) when applied to the
quadruple quasar Q2237+0305 (cf.  Table \ref{tab-cases}).
In fact, brightness fluctuations of more than one magnitude 
have been measured in Q2237+0305 on time scales
of a few months (Wo\'zniak 2000a,b), so these events do occur and
seem not to be very rare.

\subsection{Discussion and comparison with previous work}

We explored astrometric microlensing effects on short timescales
(of order months to years) for a set
      of parameter values $\kappa$ and $\gamma$.  
We find that the  relevant time scale for measuring relatively
large jumps of occasionally many tens of microarcseconds can be
as short as a few months. 
Such sudden changes of position  -- produced by caustic
crossings -- are statistically
related to fluctuations in the apparent brightness of the quasar. 
Therefore, 
a good strategy for detecting this centroid shift would be to 
measure the positions of the most promising lensed quasars
very accurately once, then 
to monitor the quasars in the optical, and      -- when a
photometric microlensing event is detected --  to perform
one or a few more accurate astrometric measurements.

We found that the effect is  most pronounced for values of
the surface mass density and shear $\kappa = \gamma = 0.4$ or 0.6. 
%{\bf 
These parameters happen to be applicable to many 
of the lensed quasar images.
%}
The most favorable case is the quadruple lens Q2237+0305:
because of the closeness of the lensing galaxy, the
time scale is relatively short (cf. also Wo\'zniak 2000a,b).
We also investigated the positional shift of the image
as a function of source size: whereas a quasar with a typical size 
of $0.16 \ \theta_E$
produces median offsets of order $ \theta_E$ (and 95th-percentiles
of about $2\ \theta_E$),     smaller sources
($0.02 \ \theta_E$) reach median values of $2.5 \ \theta_E$
and 95th-percentiles larger than $5\ \theta_E$ (where
in the case of Q2237+0305, 
$\theta_E \approx 
7.3 \times \sqrt {M /M_\odot} ~\rm  h_{75}^{-0.5} ~\mu arcsec$,
cf. Table~\ref{tab-cases}).

Lewis and Ibata (1998) had investigated the astrometric microlensing
effect specifically on Q2237+0305. 
They had found that substantial
image shifts of $\approx 100 \mu$arcsec  are possible within months.
We can   confirm this in some rare cases.
For typical caustic crossings with
photometric fluctuations of about 0.5 mag,
we find  values between 20 and  40 $\mu$arcsec.
Salata \& Zhdanov (2003) have looked
into the question of rms fluctuations of the quasar position; however,
they only considered large sources ($0.5 R_E$) and cases with 
$\kappa + \gamma < 0.8$.

%
%%%%%%%%%%%%%%%%%%%%%%%%%%%%%%%%%%%%%%%%%%%%%%%%%%%%%%%%%%%%%%%%%%%
% Figure 12
%%%%%%%%%%%%%%%%%%%%%%%%%%%%%%%%%%%%%%%%%%%%%%%%%%%%%%%%%%%%%%%%%%%
%
%
%%%%%%%%%%%%%%%%%%%%%%%%%%%%%%%%%%%%%%%%%%%%%%%%%%%%%%%%%%%%%%%%%%%
% Figure 13 
%%%%%%%%%%%%%%%%%%%%%%%%%%%%%%%%%%%%%%%%%%%%%%%%%%%%%%%%%%%%%%%%%%%
%

\subsection{Prospects for observations}

Ground-based differential astrometry in the near infrared is able to
achieve measurement uncertainties of better than 10 microarcsec, as
was      reported very recently from the Palomar Testbed Interferometer 
(Lane \& Muterspaugh 2003). This is only feasible 
 for bright objects so far,
but it is a very exciting result which opens up promising opportunities
for the coming years. 
Another instrument promising  extremely high astrometric
precision in the near future  
(with planet detection as 
one of its main scientific drivers)
is the PRIMA instrument at the ESO VLTI. It 
should become efficient in 2004 (Paresce et al. 2003;
see also Delplancke et al. 2001).
%{\bf 
The current goal at ESO is to achieve  
50 $\mu$arcsec accuracy with PRIMA 
in the $H$ and $K$ bands in 2005-2008
and 10 $\mu$arcsec accuracy in 2008-2010 (Th. Henning, private
communication).
%}

A number of space-based astrometric projects are also underway:
The  Space Interferometric Mission (SIM) is 
a five year mission 
scheduled for launch in 2009 (http://sim.jpl.nasa.gov). The
mission's goal is to reach an astrometric accuracy of 
about 1 microarcsecond for a predefined grid
of objects brighter than 13 mag in the visible,
which is quasi-inertially tied to a set of distant QSOs.
For slightly fainter objects (the four images of the lensed quasar 
Q2237+0305 have magnitudes $\sim$ 17), 
SIM is expected to yield 4 microarcsecond absolute 
positions (Unwin et al. 2002). 
In order to detect center-of-light
offsets for multiple quasars, in fact only relative astrometry between
the quasar images is required.
SIM will also be able to 
measure such positional shifts as a function of color, which
will give us hints on the physical structure of the continuum
emission region: presumably the cooler/redder part is more
extended 
than the hotter/bluer part, which means that we expect
larger changes in the center-of-light  at shorter wavelengths.
The GAIA satellite is an ESA mission currently scheduled for
launch in June 2010 (see http://sci.esa.int/gaia; Perryman et al. 2001,
Perryman 2002).
With a nominal precision of a few microarcseconds for bright
objects (about 10 microarcseconds for 15$^{th}$ mag objects), 
it will measure accurate positions of 500 000 quasars.
So GAIA  is expected to  be an extremely useful instrument
for astrometric microlensing purposes. 
It will provide many positional shifts of quasar images, 
along with their lightcurves in many filters.

\section{Summary and Conclusions}

We analyzed the shifts in the center-of-light positions
of gravitationaly lensed quasar images and the corresponding
flux variability due to the microlensing effect
of stars in the lensing galaxy. We found the following results: 

 \begin{enumerate}

 \item  
The center-of-light of multiply imaged quasars is expected to vary on
	time scales of months to years, due to 
	astrometric microlensing effects.

\item 
This effect    depends
	on the size of the lensed quasar, which we
	modelled with a Gaussian profile: 
	it is larger for smaller sources. 
	This means that 
	measuring centroid shifts will allow  us
	to constrain the size of the quasar. 

\item  
We studied eight cases, four without external shear
	($\kappa = 0.2, 0.4, 0.6, 0.8$; $\gamma = 0.0$) and
	four with shear
	($\kappa = \gamma = 0.2, 0.4, 0.6, 0.8$). The strongest
	effects are seen in the cases $\kappa = \gamma = 0.4$
	and 0.6, which in fact are very typical parameter
	values for many lensed quasar  images.

\item 
The effect of center-of-light shift of quasars is not limited
	to multiply imaged quasars: 
	single quasars may occasionally be microlensed too 
	(as was suggested by Hawkins 1996).
	The offsets should be relatively small, however, 
	due to the fact that the values of 
	$\kappa$ and $\gamma$ will be lower for single
	quasars than for multiply imaged ones.
	Our simulation with $\kappa = 0.2$ comes closest 
	to describing these cases with very low convergence.

\item 
The centroid shift effect is statistically strongly correlated with 
	changes in the apparent brightness of the quasar.
	Both occur during the crossing of a (micro-)caustic:
	photometric fluctuations $\ge 0.5$ mag produce 
		offsets 
		$\ge 2 \theta_E$ in 50\% of
		the cases, and 
		$\ge 5 \theta_E$ in 5\% of the cases.
	These values correspond to about 15 
	and 35 microarcseconds, respectively, for the
	lensed quasar Q2237+0305. 

\item 
The best strategy for measuring centroid shifts is therefore
	to monitor the
	apparent brightness of multiple quasars with ground
	based telescopes and to determine the relative
	positions of the various quasar images occasionally, 
	ideally before, during and after a 
	high magnification event.
\item 
In the cases with external shear, the center-of-light shift
	depends strongly on the direction of the shear, which
	can hence be determined in a statistical sense.

\end{enumerate}
With the next generation of astrometric instruments providing
an accuracy of order 10 microarcseconds, the 
astrometric microlensing effect of stars 
acting on background quasars will become detectable.

\vfill

\end{document}